\documentclass[a4paper]{article}
\usepackage{RRA4}
\usepackage{gastex}
\usepackage{amsmath,amssymb,verbatim,graphics,subfigure}
\usepackage[lined, algoruled]{algorithm2e} 
\usepackage[usenames]{color}
\usepackage{hyperref}

\usepackage{graphicx}
\usepackage{multicol}
\usepackage[latin1]{inputenc}
\newtheorem{Lem}{Lemma}
\newtheorem{Thm}{Theorem}

\newtheorem{Conj}{Conjecture}

\newtheorem{Ass}{Assumption}

\newenvironment{Proof}{\noindent \textbf{Proof:
}}{\hspace{\stretch{1}}$\square$}

\RRdate{December 2006}
\RRauthor{Anh-Tuan Gai\thanks{INRIA, domaine de Voluceaux, 78153 Le Chesnay cedex, France} \and Fabien Mathieu\thanks{France Telecom R\&D, 38--40, rue du g\'{e}n\'{e}ral Leclerc, 92130 Issy les Moulineaux, France} \and Julien Reynier\thanks{LIENS, 45 rue d'Ulm, 75230 Paris Cedex 05, France} \and Fabien de Montgolfier\thanks{LIAFA, 175, rue du Chevaleret, 75013 Paris France}}
\RRtitle{Stratification dans les Réseaux Pair-à-Pair\\ Application à BitTorrent}
\RRetitle{Stratification in P2P Networks\\ Application to BitTorrent}

\titlehead{Stratification in P2P networks: Application to BitTorrent}

\RRnote{CRC MARDI}
\RRabstract{
We introduce a model for decentralized networks with collaborating peers. The model is based on the stable matching theory which is applied to systems with a global ranking utility function. We consider the dynamics of peers searching for efficient collaborators and we prove that a unique stable solution exists. We prove that the system converges towards the stable solution and analyze its speed of convergence. We also study the stratification properties of the model, both when all collaborations are possible and for random possible collaborations. We present the corresponding fluid limit on the choice of collaborators in the random case.

As a practical example, we study the BitTorrent Tit-for-Tat policy. For this system, our model provides an interesting insight on peer download rates and a possible way to optimize peer strategy.}

\RRkeyword{P2P, stable marriage theory, rational choice theory, collaborative systems, BitTorrent, overlay network, matchings, graph theory}

\RRresume{Cet article vise à introduire un nouveau modèle d'analyse des réseaux décentralisés basés sur des collaborations entre pairs. Ce modèle repose sur la théorie des mariages stables appliquée à des systèmes possédant une fonction d'utilité globale. Nous étudions la dynamique induite par la recherche pour chacun des meilleurs partenaires possibles et montrons un théorème d'existence--unicité. Nous observons une rapide vitesse de convergence et étudions le phénomène de stratification dans la solution stable, dans le cas où le graphe des collaborations réalisables est complet et dans celui où il est aléatoire. Pour le cas aléatoire, nous présentons une limite fluide de la solution.

Comme exemple pratique, nous étudions la politique \emph{donnant--donnant} employée dans le logiciel de partage \emph{BitTorrent}. Pour ce système, notre modèle fournit des intuitions pertinentes sur les vitesses de téléchargements ainsi que sur les possibilités d'optimisation des paramètres.}

\RRmotcle{pair-à-pair, mariages stables, choix rationnels, systèmes collaboratifs, BitTorrent, réseaux overlay, couplages, graphes}
\RRprojets{Gyroweb}
\RRtheme{\THCom}
\URRocq 
\begin{document}
\makeRR   


\newcommand{\eme}{$^{\mathrm{\grave{e}me}}$\xspace}
\newcommand{\Ned}{\mathbb{N}}
\newcommand{\Zed}{\mathbb{Z}}
\newcommand{\norminf}[1]{\left\| #1 \right\|_\infty}
\newcommand{\normeun}[1]{\left\| #1 \right\|_1}
\newcommand{\fgauche}[1]{\overset{\lower0.5em\hbox{$\smash{\scriptscriptstyle\leftarrow}$}}{#1}}\newcommand{\un}{\textbf{1}}
\newcommand{\Red}{\mathbb{R}}
\newcommand{\bac}{\emph{Back}}
\newcommand{\abs}[1]{\left\vert#1\right\vert}
\newcommand{\set}[1]{\left\{#1\right\}}
\newcommand{\mc}[1]{\mathcal{#1}}
\newcommand{\fa}[1]{\mathcal{A}(#1)}
\newcommand{\fb}[1]{\mathcal{B}(#1)}
\newcommand{\parent}[1]{\left(#1\right)}
\newcommand{\eq}[1]{\begin{equation} #1
\end{equation}}

\newcommand{\refmark}[1]{\hbox{$^{\ref{#1}}$}}
\newcommand{\fref}[1]{\ifthenelse{\equal{\pageref{#1}}{\thepage}}%
{\ref{#1}}%
{\ref{#1} pa\-ge~\pageref{#1}}}

\section{Introduction}

\paragraph{Motivation} 
Collaboration-based distributed applications are successfully applied to large scale systems. A system is said to be collaborative when participating peers collaborate in order to reach their own goal (including being altruistic). Apart from well-known content distribution applications \cite{cohen03incentives, edonkey}, collaborating can be applied to numerous applications such as distributed computing, online gaming, or cooperative backup. The common property of such systems is that participating peer exchange resources. The underlying mechanism provided by protocols for such applications consists in selecting which peers to collaborate with to maximize one's peer benefit with regards to its personal interest. This mechanism generally uses a utility function taking local information as input. One can ask if this approach can provide desirable properties of collaboration-based content distribution protocols like scalability and reliability.

To achieve these properties, the famous protocol BitTorrent \cite{cohen03incentives} implements a Tit-for-Tat (TFT) exchange policy. More precisely, each node \emph{knows} a subset of all other nodes of the system and collaborates with the best ones from its point of view: it uploads to the contacts it has most downloaded from in the last 10 seconds. In other words, the utility of peer $p$ for node $q$ is equal to the quantity of data peer $q$ has downloaded from $p$ (in the last measurement period). The main interest in using the TFT policy is the resulting incentive to cooperate. The nature of the utility function then leads to a clustering process which gather peers with similar upload performances together, called \emph{stratification}.

Recently, much research has been devoted to the study of the phenomenon. So far, however, while it has been measured and observed by simulations, it has not been formally proved. Understanding stratification is a first step towards a better comprehension of the impact of the utility function on a system behavior. 
A theoretical framework to analyze and compare different utility function is needed: choosing a utility function that best suits a given application is quite difficult. More importantly, it is not clear whether the utility functions implemented lead to desirable properties. We introduce a generic framework that allows an instantiation of (known and novel) utility functions that model collaboration. We further present a thorough analysis of a class of utility functions based on global ranking agreements, such as that of BitTorrent TFT policy. This framework also fits gossip-based protocols used by a peer to discover its rank~\cite{gossip}.

\paragraph{Contribution}
First, we propose a model based on the stable matching theory. This model describes decentralized networks where peers rank each others and try to collaborate with the best peers for them.

Second, we focus on systems with a global ranking utility function (each peer has an intrinsic value) in the framework of stable matching. We prove that such a system always admits a unique stable solution towards which it converges. We verify through simulations the speed of convergence without and with \emph{churn} (arrivals ans departures).

Third, we study stratification in a toy model of  fully connected networks where every peer can collaborate with all other peers. If every peer tries to collaborate with the same number of peers, we observe disjoint clustering. But with a variable number of collaborations per peer, clustering turns into strong stratification.

Fourth, we describe stratification in random graphs. For Erd\"{o}s-R\'{e}nyi graphs, the distribution of collaborating peers has a fluid limit. This limiting distribution shows that stratification is a scalable result.

Lastly, we propose a practical application of our results to the BitTorrent TFT policy. Assuming content availability is not a bottleneck in a BitTorrent swarm, our model leads to an interesting characterization of the download rate a peer can expect as a function of its upload rate. This description leads to possible strategies for optimizing the download for a given upload rate.

\paragraph{Roadmap}
In Section~\ref{sec:context} we define our model. Section~\ref{sec:ranking} presents a study on the problem dynamics. Section~\ref{sec:complete} describes stratification in a complete neighborhood graph and Section~\ref{sec:global_random}, in random graphs. Section~\ref{sec:consequences} discusses the application of our results to BitTorrent and Section~\ref{sec:conclusion} concludes the paper.

\section{Model}
\label{sec:context}

P2P networks are formed by establishing an overlay network between peers.
A peer acts both as a server and a client. Each peer $p$ has a bounded number $b(p)$ of collaboration slots. As the network evolves, peers continuously search after new (or better) partners. 
Each protocol has its own approach to handling these dynamic changes. For example, a protocol like eDonkey~\cite{edonkey,emule} optimizes independently two preference lists on the server and on the client sides. More recent protocols, like BitTorrent ~\cite{cohen03incentives}, make a use of a game theoretic approach, where each peer tries to improve its own payoff. It results in keeping one preference list per node.

Let us suppose that each peer $p$ has a \emph{global mark} $S(p)$, which may represent its available bandwidth, its computational capacities, or its shared storage capacity. Each peer wants to collaborate with best partners who have highest marks $S(p)$. 
This models many networks preferences
systems, albeit not all networks have such ranking. For instance, in
chess playing, players have an intrinsic value (ELO rating), although
they don't generally want to engage people far better or worse than
them.

Some peers might not be willing to cooperate with some others. For instance, peers that have no common interest or are unaware of each other.
We introduce an \emph{acceptance graph} to represent compatibilities. A pair $(p,q)$ belongs to the acceptance graph if, and only if (iff) both peers are interested in collaboration. Without loss of generality, we can suppose acceptability is a symmetric relation: if $p$ is unacceptable for $q$, $q$ will never be able to collaborate with $p$ so we can assume $q$ is also unacceptable for $p$. 
We denote by \emph{configuration} or \emph{matching} the subgraph of the acceptance graph that represents the effective collaboration between peers. The degree of a peer $p$ in a configuration is bounded by $b(p)$.

A \emph{blocking pair} for a given configuration is a set of two peers unmatched together wishing to be matched together (even if it means dropping one of their current collaborations). A configuration without blocking pair is said to be \emph{stable}. In a stable configuration, a single peer cannot improve its situation: it is a Nash equilibrium.

If a number of colloborations is limited to $1$, the problem is known as the \emph{stable roommates problem} \cite{irving87efficient}. It is an extension of the famous \emph{stable marriage problem} introduced by Gale and Shapley in 1962 \cite{gale62college}. If we assume each peer $p$ wants to collaborate with up to $b(p)$ other peers, the framework is called stable $b$-matching problem\footnote{in this paper the word \emph{matching} stands for $b$-matching (unless otherwise stated)}
\cite{fleiner05generalization}.

As it holds for all theories of stable matchings, the existence of a stable configuration depends on the preference rules used to rank participant and on the acceptance graph. In this work we study the impact of the rules derived from a global ranking on a peer-to-peer network behavior. In particular, we find the properties of the stable configurations.

\section{Existence and convergence properties of a stable configuration}
\label{sec:ranking}

Global ranking matching is one of the simplest cases of matching problems. Tan \cite{tan91necessary} has shown that existence and uniqueness of
stable solutions were related to preference cycles in the utility function. A preference cycle of length $k$ is a set $i_1,\ldots,i_k$ of $k$
distinct peers such that each peer of the cycle prefers its successor to its
predecessor. As proved by Tan, a stable configuration exists iff
 there is no odd preference cycle of length greater than $1$. He also proved
that if no even cycle of length greater than $2$ exists, then the stable configuration is unique. If peers 
have an intrinsic value, no strict preferences cycle can occur (see below for ties), so a global ranking matching problem has one and only one stable solution.

This solution is very easy to compute knowing the global ranking $S$, $b$ and the acceptance graph. The process is given by Algorithm~\ref{alg:stable}: each peer $p$ starts with $b(p)$ available connections. First, the best peer $p_1$ picks the best $b(p_1)$ peers from its acceptance list. As $p_1$ is the best, the chosen peers gladly accept (recall the acceptance graph is symmetric) and the resulting collaborations are stable (no blocking pair can unmatch them). Note that if there is not enough acceptable peers, $p_1$ may not satisfy all its connections. Peers chosen by $p_1$ have one less connection available. Then second best peer $p_2$ does the same, and so on\ldots By immediate recurrence, all connections made are stable. 
 When the process reaches the last peer, the connections are the stable configuration for the problem. 
As it was said before, all connections are not necessarily satisfied. For instance, if the last peer still has available connections when its turn comes, his connections will not be fetched, as all peers above him have by construction spent all their connections. This is, of course, a centralized algorithm, but we shall see below that decentralized algorithms work as well.

\begin{algorithm}
\setlength{\baselineskip}{0.6\baselineskip} 
\dontprintsemicolon
\SetKwData{avc}{cur\_b}
\SetKwFunction{connect}{connect}
\KwData{
An acceptance graph $G$ with $n$ peers, a global ranking $S(p)$, and maximal number of connections $b(p)$ }
\KwResult{The unique stable configuration of the $b$-matching problem\;}
\BlankLine
Let $a$ be a vector initialized with $b$\;
\For{each peer $i$ sorted in increasing $S(p)$ (best peer first)}{
\For{each peer $j$ sorted in increasing $S(p)$ starting just after $i$}{
\If{$(i,j)\in G$  and $a(i)>0$ and $a(j)>0$}{
\connect{$i,j$}\;
$a(i)=a(i)-1$\;
$a(j)=a(j)-1$\;
}}}
\caption{Stable configuration in global ranking}
\label{alg:stable}
\end{algorithm}

\paragraph*{Note on ties}
Ties in preference lists make the matching problems more difficult to resolve \cite{ronn86complexity} without bringing more insight about the stratification issues studied in this paper. Simulations have shown our results hold if we allow ties, but equations are hard to prove as existence of
a stable matching cannot be guaranteed. Thus for the sake of simplicity, we shall suppose utilities are distinct, that is $S(q)\neq S(p)$ for any
$p\neq q$.

\paragraph*{Convergence}
One can ask what is the point in studying a stable configuration in a dynamical context such as P2P systems, where peers arrive and depart whenever they 
wish, and where utility functions and acceptance lists can fluctuate. We have not proved yet that the process of peers trying independently to collaborate to the best peers they know can reach the stable state.

We introduce the concept of \emph{initiative} to model the process by which a peer may change its mates. Given a configuration $C$, we say that peer $p$ \emph{takes the initiative} when it proposes to other peers to be its new mate.
Basically, $p$ may propose partnership to any acceptable peer. But only blocking pairs of $C$ represent an interesting new partnership.  If $p$ can find such a \emph{blocking mate}, the initiative is called \emph{active} because it succeeds in modifying the configuration (both peers will change their set of mates). 

To find a blocking mate, $p$ contacts peers from its acceptance list.
We identify several strategies depending on how $p$ scans its acceptance list:
\begin{description}
\item[best mate] when the peer selects the best (if any) available blocking mate. This happens if $p$ knows the rank of all its acceptable peers and whether they will collaborate or not,
\item[decremental] when the list is circularly scanned starting from the last asked peer. This happens if $p$ knows the rank of all its acceptable peers, but not if they will collaborate,
\item[random] when a single peer is selected at random. This happens if $p$ has no information on its neighbors until it asks.
\end{description}

Of course, when best mate initiative is possible, it seems to be the best strategy to maximize a peer's own profit, but it supposes a good knowledge of the system is maintained.

We can now complete our model with initiatives: starting from any initial configuration, an instance of our model evolves because of initiatives taken by peers. In fact, it can only evolve towards the unique stable configuration, as shown by Theorem~\ref{thm:acyclic}.

\begin{Thm}
\label{thm:acyclic} The stable solution can be reached in $B/2$ initiatives, where $B=\sum_{p}b(p)$ is the maximal number of connections. Moreover, any sequence of active initiatives starting from any initial configuration eventually reaches the stable configuration.
\end{Thm}

\begin{Proof}
In Algorithm~\ref{alg:stable}, each connection can be obtained by initiative. As the stable configuration possesses up to $B/2$ pairings, this ensures the first part of the theorem. We prove the convergence by showing a sequence of active initiatives can never produce twice the same configuration. There is a finite number of possible configurations, so if we keep altering the configuration through initiatives, we eventually reach a configuration that cannot be altered with any initiative: the stable configuration.

The proof is indeed simple. If a sequence of initiatives induces a cycle of at least two distinct configurations, then one can extract a preference cycle of length greater than $3$: let $p_1$ be a peer whose mates change through the cycle. Call $p_2$ the best peer $p_1$ is unstably paired with during the cycle, and  $p_3$ the best peer $p_2$ is unstably paired with during the cycle. $p_1$ is not $p_3$ and $p_2$ prefers $p_3$ to $p_1$, otherwise the pair $\{p_1,p_2\}$ would not break during the cycle. Iterating the process, we build a sequence of peer $(p_k)$ such that $p_k$ prefers $p_{k+1}$ to $p_{k-1}$, until we find $i<j$ such that $p_i=p_j$. The circular list $(p_i,p_{i+1},\ldots,p_{j-1})$ is a preference cycle. As global ranking does not allow preference cycles, this is not possible, so a sequence of active initiatives can never produce twice the same configuration.
\end{Proof}

Theorem \ref{thm:acyclic} proves that in static conditions (no join or departure, constant utility function), a P2P system will converge to the
stable state. To prove this stable state is worth studying, we have to show convergence is fast in practice (Algorithm~\ref{alg:stable} is optimal in number of initiatives but difficult to implement in a large scale system) and can sustain a certain amount of churn. As a complete formal proof of this is beyond the scope of this paper, we used simulations.

In our simulations, peers were labeled from $1$ to $n$ (the number of peers). These labels define the global ranking, $1$ being the best peer and $n$ the worst (if $i<j$, peer $i$ is better than peer $j$). We use Erdös-Renyi loopless symmetric graphs $G(n,d)$ as acceptance graphs, where $d$ is the expected degree (each edge exists independently with probability $\frac{d}{n-1}$). Only $1$-matching was considered.

For measuring the difference between two configurations $C_1$ and $C_2$ we use the distance

$$D(C_1,C_2)=\Sigma_{i=1}^n\|\sigma(C_1,i)-\sigma(C_2,i)\|.\frac{2}{n(n+1)}\text{,}$$

\noindent where $\sigma(C,i)$ denotes the mate of $i$ in $C$ (by convention, $\sigma(C,i)=n+1$ if $i$ is unmated in $C$).

$D$ is normalized: the distance between a complete matching and the empty configuration $C_\emptyset$ is equal to $1$. The \emph{disorder} denotes the distance between the current configuration and the stable configuration.

At each step of the process we simulate, a peer is chosen at random and performs a best mate initiative (the initiative can be active or not). To compare simulations with different number $n$ of peers, we take a sequence of $n$ successive initiatives as a base unit (that can be seen as \emph{one expected initiative per peer}).

\begin{figure}
\begin{center}
\includegraphics[width=.7\textwidth]{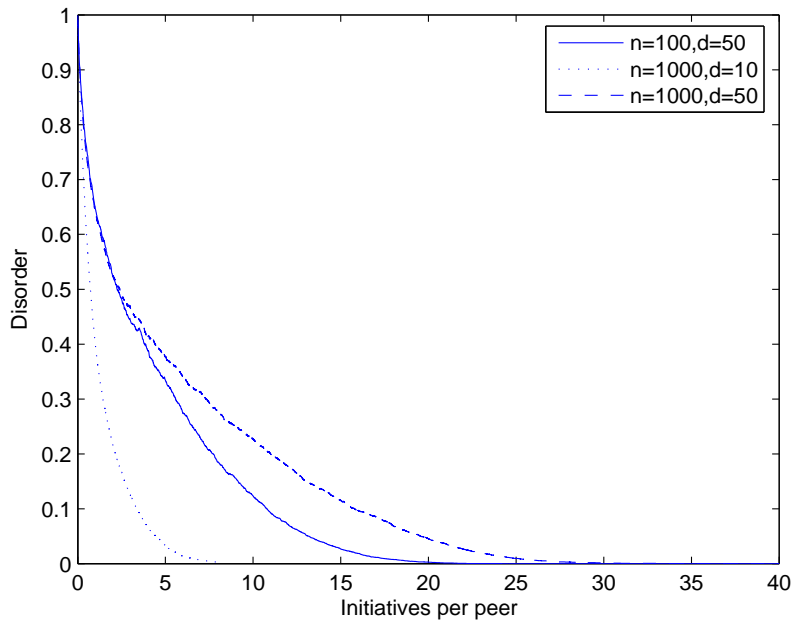}
\end{center}
\caption{Starting from $C_\emptyset$, convergence towards the stable state for different parameters}
\label{fig:nochurn}
\end{figure}

A first set of simulations is made to prove a rapid convergence when the acceptance graph is static. 
In all simulations, the disorder quickly decreases, and the stable configuration is reached in less than $nd$ initiatives (that is $d$ base unit). Figure~\ref{fig:nochurn} shows convergence starting from the empty configuration for three typical parameters: $(n,d)=(100,50)$, $(n,d)=(1000,10)$, $(n,d)=(1000,50)$.

\begin{figure}
\begin{center}
\includegraphics[width=.7\textwidth]{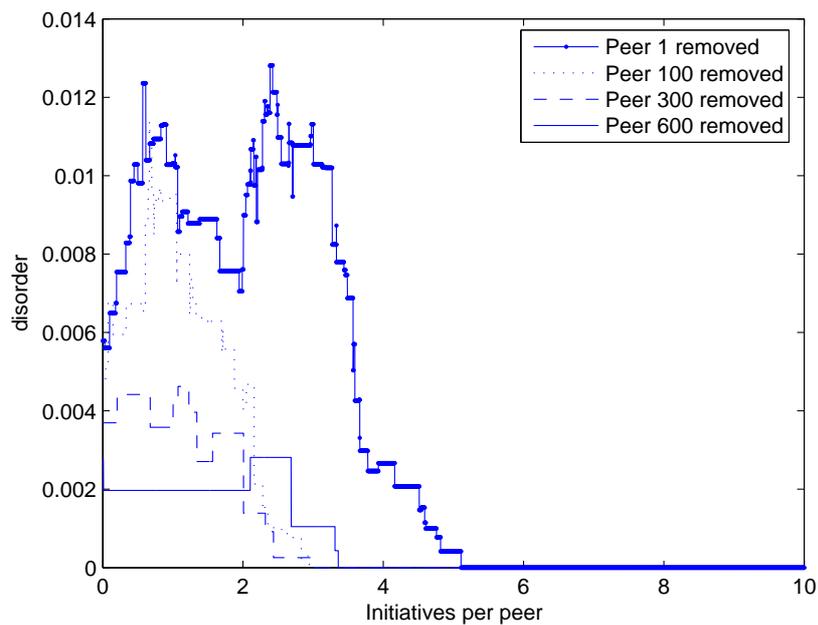}
\end{center}
\caption{Starting from the stable state, we remove a peer and observe the convergence towards the new stable state. ($1000$ users, $1$-matching, $10$ neighbors per peer)}
\label{fig:healing}
\end{figure}

Then we investigate the impact of an atomic alteration of the system.
 Starting from the stable configuration, we remove a peer from the system and observe the convergence towards the new stable configuration. We observe big variances in convergence patterns, but convergence always takes less than $d$ base units and disorder is always small. Note, that due to a domino effect, removing a good peer generally induces more disorder than removing a bad peer. This is shown by Figure~\ref{fig:healing}. We ran the simulations $100$ times and selected four representative trajectories, as we did not wish to average out interesting patterns. 

\begin{figure}
\begin{center}
\includegraphics[width=.7\textwidth]{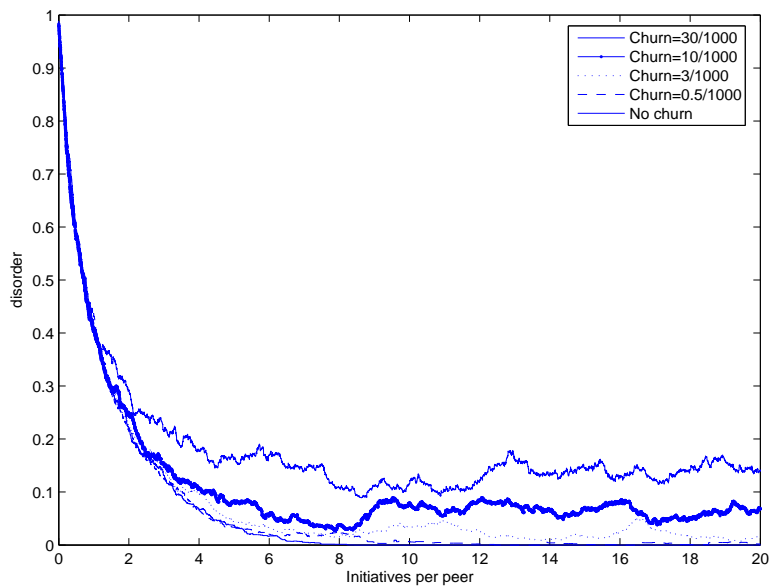}
\end{center}
\caption{Starting from $C_\emptyset$, we observe distance to the instant stable state with different churn levels ($1000$ users, $1$-matching, $10$ neighbors per peer)}
\label{fig:churn}
\end{figure}

Finally, we investigate continuous churn. A peer can be removed or introduced in the system anytime, according to a \emph{churn rate} parameter. Simulations show that as the churn rate increases, the system becomes unable to reach the instant stable configuration. However, the disorder is kept under control. That means the current configuration is never far from the instant stable configuration. The average disorder is roughly proportional to the churn rate (see Figure~\ref{fig:churn} for typical patterns).

All these simulations lead to the same idea: the stable configuration acts like a strong attractor in the space of possible configurations when collaborations are established using intrinsic values for judging peers Studying the properties of the stable configuration is the next step.

\section{Stratification with complete acceptance graph}
\label{sec:complete}

We start studying the stable configuration in the special case where everybody is acceptable for everybody. Hence the acceptance graph is complete. This is a suitable, but not scalable, assumption for small systems. Complete acceptance graph is a toy model for highlighting stratification effect.

\subsection{Clustering in constant \texorpdfstring{$b$}{b}-matching}

\begin{figure*}[htb]
\begin{center}
\begin{picture}(136,19)(14,-55)\nullfont
\node(n0)(16.0,-44.0){$1$}

\node(n1)(28.0,-44.0){$2$}

\node(n2)(40.0,-44.0){$3$}

\node(n3)(52.0,-44.0){$4$}

\node(n4)(64.0,-44.0){$5$}

\node(n5)(76.0,-44.0){$6$}

\node[Nadjust=w](n6)(104.0,-44.0){$3k+1$}

\node[Nadjust=w](n7)(116.0,-44.0){$3k+2$}

\node[Nadjust=w](n8)(128.0,-44.0){$3k+3$}

\drawedge[AHnb=0,curvedepth=-8.0](n0,n1){}

\drawedge[AHnb=0,curvedepth=-8.0](n1,n2){}

\drawedge[AHnb=0,curvedepth=-8.0](n3,n4){}

\drawedge[AHnb=0,curvedepth=-8.0](n4,n5){}

\drawedge[AHnb=0,curvedepth=-8.0](n6,n7){}

\drawedge[AHnb=0,curvedepth=-8.0](n7,n8){}

\drawedge[AHnb=0,curvedepth=-7](n2,n0){}

\drawedge[AHnb=0,curvedepth=-7](n5,n3){}

\drawedge[AHnb=0,curvedepth=-7](n8,n6){}

\drawline[AHnb=0,dash={1.0 1.0 1.0 1.0}{0.0}](82,-44)(96,-44)

\drawline[AHnb=0,dash={1.0 1.0 1.0 1.0}{0.0}](134,-44)(150,-44)





\end{picture}
\caption{Limit case of $b$-matching and total knowledge : the collaboration graph is a set of $b+1$ clusters. Here $b=2$.}
\label{fig:cluster}
\end{center}
\end{figure*}
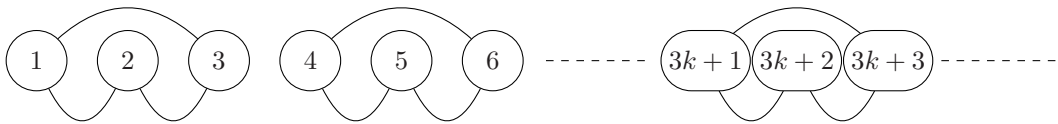

Constant $b$-matching is an instance of the $b$-matching problem where every one tries to connect to at most $b_0$ peers ($b_0$ is a constant). Since the acceptance graph is complete, the stable configuration is very simple. It consists in a sequence of complete subgraphs with $b_0+1$ elements starting from the best peer (the remainder, if any, is a truncated complete subgraph). For example, figure \ref{fig:cluster} shows this \emph{clustering} for the $2$-matching problem on a complete graph.

As it has already been pointed out \cite{bharambe06analyzing}, full clustering in file sharing networks induces poor performances. Many designers try
to produce overlay graphs with small world properties: almost fully connected, high clustering coefficient,
low mean distance, and navigable such that shortest paths may be greedily found. But in file sharing networks, having a compliant overlay with nice properties (connectivity, distances, resilience) is useless if the effective collaborations graph has none of the desired properties. In our example, although the knowledge graph is a complete graph, collaboration established through global ranking scatters the graph in clusters. Hence content is sealed inside clusters, and singularities are bound to occur.

\subsubsection*{Lower bound for number of slots in BitTorrent}
As we have just spoken of clustering, it is interesting to remember that a connected graph of $n$ vertex has at least $n-1$ edges. As a $b_0$-regular graph has  $\frac{b_0}{2}n$ edges, it is impossible for a  $1$-regular graph to be connected, and the cycle is the unique 2-regular connected graph. It follows that it is better to set $b_0\ge 3$.

This gives a first basic insight for the fact that the default number of slots per user is $4$ is BT (less for very small connections and more for high
bandwidth ones): given the generous extra slot, put less than $4$ slots in the default client would make the TFT collaboration graph disconnected
which would seriously harm the BT efficiency.

Of course, BT is more complicated, and this is just a by-passing remark. In Section \ref{sec:consequences} we propose further arguments to see why $4$ seems to be the number of connections the average client should set by default.

\subsection{Stratification in variable \texorpdfstring{$b$}{b}-matching}

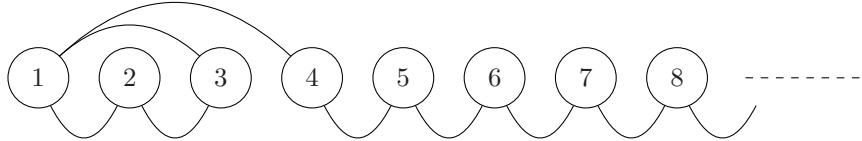
\begin{figure*}
\begin{center}
\begin{picture}(110,19)(14,-55)\nullfont
\node(n0)(16.0,-44.0){$1$}

\node(n1)(28.0,-44.0){$2$}

\node(n2)(40.0,-44.0){$3$}

\node(n3)(52.0,-44.0){$4$}

\node(n4)(64.0,-44.0){$5$}

\node(n5)(76.0,-44.0){$6$}

\node(n6)(88.0,-44.0){$7$}

\node(n7)(100.0,-44.0){$8$}

\node[Nframe=n](n8)(112.0,-44.0){}

\drawedge[AHnb=0,curvedepth=-8.0](n0,n1){}

\drawedge[AHnb=0,curvedepth=-8.0](n1,n2){}

\drawedge[AHnb=0,curvedepth=-8.0](n3,n4){}

\drawedge[AHnb=0,curvedepth=-8.0](n4,n5){}

\drawedge[AHnb=0,curvedepth=-8.0](n6,n7){}

\drawedge[AHnb=0,curvedepth=-8.0](n7,n8){}

\drawedge[AHnb=0,curvedepth=-7.0](n2,n0){}

\drawline[dash={1.0 1.0 1.0 1.0}{0.0},AHnb=0](108,-44)(124,-44)

\drawedge[AHnb=0,curvedepth=10](n0,n3){}

\drawedge[AHnb=0,curvedepth=-8.0](n5,n6){}





\end{picture}
\caption{$b$-matching plus one extra connection: the graph is connected}
\label{fig:connected}
\end{center}
\end{figure*}

$b_0$-matching is not the most common case in practice. The clustering from Figure \ref{fig:cluster} may be a consequence of the specific parameters used. Indeed, adding only one connection can alter a set of complete subgraphs of size $b_0+1$ in one unique connected component (see Figure \ref{fig:connected} -- settings
are same than for Figure \ref{fig:cluster} except that an extra connexion has been granted to peer $1$).

\begin{figure}[ht]
\centering
\includegraphics[width=.7\textwidth]{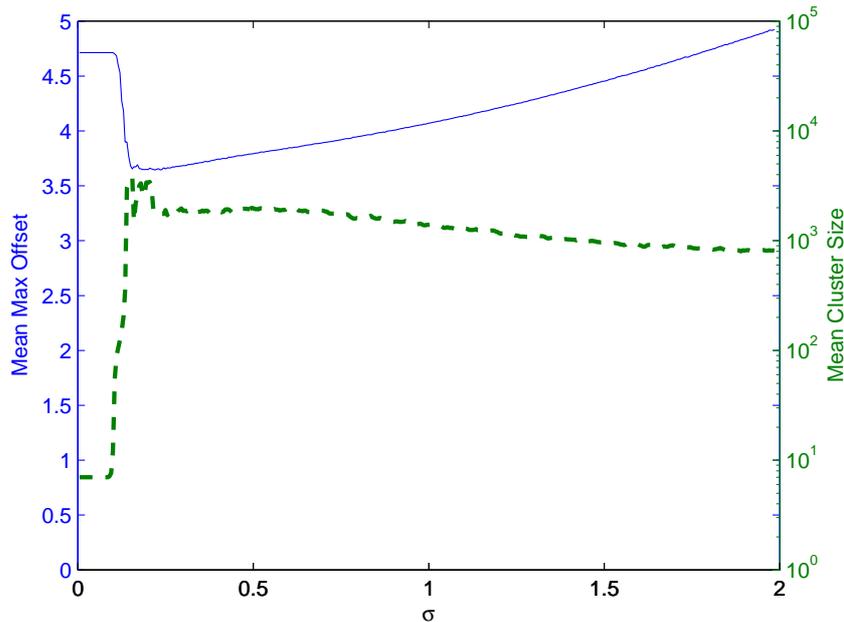}
\caption[]{Influence of $\sigma$ for the $b$-matching global ranking problem when $b$ follow a normal law ${\cal N}(6,\sigma)$. The dotted line represents the Mean Cluster Size, the plain line stands
for Mean Max Offset. The left part (when $\sigma=0$) is the case of constant 6-matching} \label{fig:sixmatching}
\end{figure}

In fact, both Figures \ref{fig:cluster} and \ref{fig:connected} are not typical. In our simulations on complete acceptance graphs, we generally observed many large connected components. If we assume that $b$ is distributed according to a rounded normal distribution ${\cal N}(\bar{b},\sigma^2)$
(mean $\bar{b}$, variance $\sigma$, all samples are rounded to the nearest positive integer), we observe a surprising phase transition. As soon $\sigma$ is big enough to produce heterogeneous samples ($\sigma \approx 0.15$), the average connected component size explodes, then stays almost constant. The cluster typical size after the transition seems to grow factorially with $\bar{b}$ (Figure \ref{fig:sixmatching} shows what happens for $\bar{b}=6$).
Computed values appear in Table \ref{tab:complete}.

\begin{table}
    \centering
        \begin{tabular}{|r||c|c|c|c|c|c||c|c|c|c|c|c|}
\cline{2-13}
\multicolumn{1}{c||}{} & \multicolumn{6}{c||}{constant $b_0$-matching} &  \multicolumn{6}{c|}{normal  ${\cal N}(\bar{b},\sigma^2)$-matching with $\sigma=0.2$ }\\
        \hline
  $b_0$ or $\bar{b}$            & $2$ & $3$ & $4$ & $5$ & $6$ & $7$ &  $2$ & $3$ & $4$ & $5$ & $6$ & $7$\\
            \hline
            \textbf{Average Cluster Size } &$3$&$4$&$5$&$6$&$7$&$8$  & 6 & 20 & 78 & 350 & 1800 & 11000\\
            \hline
            \textbf{Max Mean Offset (MMO)} & 1.67 & 2.5 & 3.2 & 4 & 4.71 & 5.5 & 1.33&2.10&2.52&3.21&3.65&4.31\\
            \hline
            \end{tabular}
\caption{Clustering and stratification properties in a complete knowledge graph. }
\label{tab:complete}
\end{table}

Factorial cluster size growth grants the existence of a giant connected component when $\bar{b}$ is large and $n$ remains bounded. This solves the
clustering issue.

Nevertheless, distances in the obtained collaboration graph are another question. A good estimate is given by Mean Max Offset (MMO) which described
the mean ranking offset between one peer and its further neighbor in the collaboration graph. The larger the MMO, the fewer hops needed to link two
peers with very different intrinsic value in the same connected component. Remark that in $b_0$-matching, MMO is easy to compute (it is enough to compute it on the $b_0+1$ complete
graph). We show that it converges to:

\begin{eqnarray*}
MMO(b_0) & = \frac{1}{b_0+1}(b_0+(b_0-1)+\ldots+\left\lceil \frac{b_0}{2}\right\rceil+\ldots+b_0)&\\
& \xrightarrow[b_0\rightarrow +\infty]{} \frac{3}{4}b_0.
\end{eqnarray*}

When $b$ is variable, MMO becomes less obvious to compute. However, simulations show that MMO reflects the same phase transition as the cluster size does. In contrast, as cluster size explodes, MMO decreases, has shown by  Figure \ref{fig:sixmatching} and Table \ref{tab:complete}.

The conclusion of this first approach on complete graphs is that whereas the clustering problem can be handled, a stratification issue exists: peers
only collaborate with very close to them, which can make content diffusion ineffective.

\section{Global ranking on random acceptance graphs}
\label{sec:global_random}

In this section, for the sake of simplicity, we first describe a $1$-matching model. This allows us to explain our {\it independence} assumption and to present the related mathematical results. We then extend the equations to the $b_0$-matching case, for any constant number $b_0$ of connexions.

\subsection{Model}
\label{sec:model}
As noted in Section \ref{sec:ranking}, there exists a unique matching (stable configuration) where no peer can locally improve its mates among
its known peers; this matching, denoted $P$ in the following, can be obtained simply by applying algorithm \ref{alg:stable}. As we first focus on 1-matching, we denote $P(i)$ the mate of Peer $i$ in $P$.

\subsubsection{An exact formula}
Denote by $D(i,j)$ the probability that Peer $i$ is matched with Peer $j$ over all possible graphs with $n$ vertices. In other words $D(i,.)$ is
the distribution of the peer matched with $i$.

Obviously, $D(i,j)=D(j,i)$ and $D(i,i)=0$. The total order property can be written as follows, for $i<j$:

$D(i,j)=p\mathbb{P}($"$i$ is not with better than $j$" and "$j$ is not with better than $i$"$)$ where $p$ is the probability
that peer $i$ knows peer $j$.

We can rewrite the above probability as: $\mathbb{P}(P(i)\geq j)\times \mathbb{P}(P(j)\geq i| P(i)\geq j)$, leading to the exact formula:
\begin{equation}\label{RecurrenceD_exact}
D(i,j)=p(1-\sum_{k=1}^{j-1}D(i,k))\mathbb{P}(P(j)\geq i| P(i)\geq j).
\end{equation}
Note that this does not depend on the number of peers. The formula can thus be extended to every couple $(i,j)\in\left(\mathbb{N}^*\right)^2$.

\begin{Lem}\label{Lem:Exact_mass_1}
$$\forall i\in\mathbb{N}^*, \sum_{k=1}^{\infty}D(i,k)=1.$$
\end{Lem}
This lemma means that, under the Erd\"{o}s-R\'{e}nyi assumption, when adding a large number of peers at a lower rank, any peer will eventually find a mate with probability one.

\begin{Proof} { \bf The conditional probability does not go to $0$:}
We first show that $\mathbb{P}(P(j)\geq i| P(i)\geq j)$ does not go to $0$. Suppose that $j>i$, then condition on $E_i=\{P(1),...,P(i-1)\}$,
\begin{equation*}
\mathbb{P}(P(j)\geq i| P(i)\geq j|E_i)=
\\
\left\{\begin{array}{c}
   \hbox{empty conditioning if } i\in E_i\\
    0 \hbox{ if\ } j \in E_i \hbox{(and }i \notin E_i\hbox{)}\\
    x\geq p \hbox{ if\ } j\notin E_i \hbox{ and }i \notin E_i.\\
\end{array}
\right.
\end{equation*}

The last inequality holds because if $j\notin E_i$ and $i \notin E_i$, then knowing that $P(i)\geq j$, $i$ and $j$ are linked if and only if there
exists an edge between both. Since $P(j)=i$ implies as a particular consequence $P(j)\geq i$, the inequality is satisfied.

Now, all we have to show is that $\mathbb{P}(j\in E_i|i\notin E_i)$ does not tend to $1$ when $j$ tends to infinity. This is obvious since for some
$k<i$, the function $j\to \mathbb{P}(P(k)=j|i\notin E_i)$ gives probabilities of disjoint events so that $\sum_{j=1}^{\infty}\mathbb{P}(j\in
E_i|i\notin E_i)\leq i-1$; the general term thus tends to $0$ and certainly not to $1$.

\paragraph{$D$ is a probability}
We know that for a given $i$, $D(i,j)$ are the probabilities of disjoint events. Thus $D(i,j)\xrightarrow[j\to\infty]{} 0.$ From formula
(\ref{RecurrenceD_exact}) we deduce $$\sum_{k=1}^{j-1}D(i,k)\xrightarrow[j\to\infty]{} 1.$$

\end{Proof}

\subsubsection{Approximation: independent \texorpdfstring{$1$}{1}-matching model}
Hereinafter we shall adopt the following assumption:
\begin{Ass}\label{AssHole}
the two events:
\begin{itemize}
\item peer $i$ is not with a peer better than $j$,
\item peer $j$ is not with a peer better than $i$,
\end{itemize}
are independent.
\end{Ass}
Assumption \ref{AssHole} is reasonable when the probability that $i$ and $j$ have a common neighbor is very low. It entails that (\ref{RecurrenceD_exact}) can be replaced by the approximate recurrence relation:

\begin{equation}
\label{RecurrenceD} D(i,j)=p\left(1-\sum_{k=1}^{j-1} D(i,k)\right)\left(1-\sum_{k=1}^{i-1} D(j,k)\right)
\end{equation}

This formula can easily be computed in an iterative way by calculating for increasing $i$ the probabilities $D(i,j)$ from $j=1$ to $n$ using Algorithm~\ref{alg:independent1} (see Algorithm~\ref{alg:independentb} for the $b_0$-matching case).

\begin{algorithm}\setlength{\baselineskip}{0.6\baselineskip} 
\dontprintsemicolon \SetKwData{avc}{cur\_b} \SetKwFunction{connect}{connect} \KwData{ Number of peers, $n$\; Erd\"{o}s-R\'{e}nyi probability, $p$\;}
\KwResult{$D(i,j)$ the probability user $i$ chooses user $j$\;}
 \BlankLine
$D\leftarrow zeros(n,n)$\;  \For{$i = 1$ \emph{\KwTo} $n$} { \For{$j = i+1$ \emph{\KwTo} $n$}{ $D(i,j)\leftarrow
p\left(1-\sum_{k=1}^{j-1}D(i,k)\right) \left(1-\sum_{k=1}^{i-1}D(j,k)\right)$\; $D(j,i)\leftarrow D(i,j)$ \; }} \caption{Independent $1$-matching probability computation} \label{alg:independent1}
\end{algorithm}

\begin{figure}
\includegraphics[width=\textwidth]{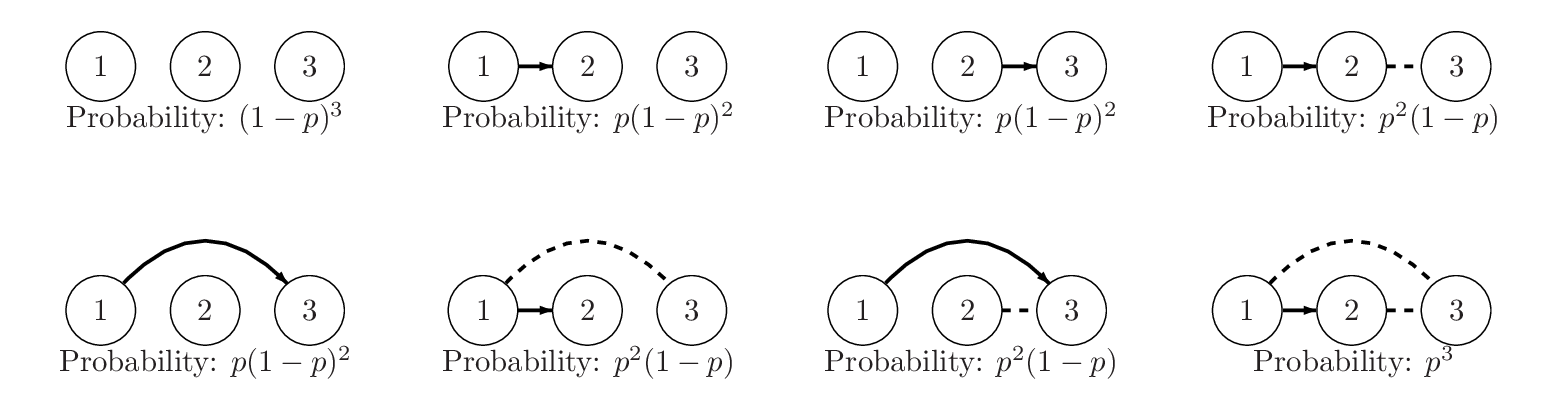}
\caption[A counter-example for $n=3$]{Approximation error: for $n=3$, there is $8$ possible graphs. Exact matchings probabilities are\\
-- $D_{\text{exact}}(1,2)=p$\\
-- $D_{\text{exact}}(1,3)=p(1-p)$\\
-- $D_{\text{exact}}(2,3)=p(1-p)^2$\\
Algorithm \ref{alg:independent1} leads to the same except\\
$\begin{array}{rcl}
D(2,3) & = & p(1-D(2,1))(1-D(3,1))\\
& = & p(1-p)(1-p(1-p))\\
&=&D_{\text{exact}}(2,3)+p^3(1-p)
\end{array}$}
\label{fig:contrex}
\end{figure}

\paragraph*{Example where the simplified formula does not work}
Even if the approximation made by using \eqref{RecurrenceD} instead of \eqref{RecurrenceD_exact} works very well for small values of $p$ (see figure
\ref{fig:3000sur5000_D1_D2_SimulatedvsEstimated}), it in not an exact formula. Example in Figure \ref{fig:contrex} illustrates this point: we
considered $3$ peers; then we can write down all the possible graphs ($8$ of them) with the exact probability for each event.

\subsection{Main result on the independent \texorpdfstring{$1$}{1}-matching model}
\label{sec:main} This section presents mathematical results that follow from assumption \ref{AssHole}. When the number of peers is large, the model scales and the normalized histogram of neighbors tends to a continuous distribution and yields
an equation satisfied in this limit. 
Indeed the empirical distribution also converges, which means that every instance of an
Erd\"{o}s-R\'{e}nyi graph is very likely to behave like the typical case of the above assumption, as shown by the simulations below.

We are able to prove some parts of this program but must leave the remainder as conjectures for further work. The results bring considerable insight.

From a practical point of view one only need to retain two points from the mathematical developments:
\begin{itemize}
 \item for not moderate values of $n$, there exists a scaled version of $D(i,j)$ which does not depend on $n$ (see \ref{subsec:observe}),
 \item the shape of $D(i,j)$ is present in almost any given $n$-peers system.
\end{itemize}

\subsubsection{Distribution weak convergence}
 {\sc Notation, Hypotheses: } For all theorems and proofs of this sections, $G=(V,E)$ is an Erd\"{o}s-R\'{e}nyi graph,  $P$ is its unique
stable pairing, and ${\cal M}_i(n,p)$ is the distribution of the mate of peer $i$:
$${\cal M}_i(n,p)=\sum_{j\in[|1,n|]\backslash \{i\}} D(i,j) \delta_{j}.$$
The mean degree of a peer is denoted $d$.

\begin{Thm}\label{th1} ${\cal M}_i(n,p)\xrightarrow[n\to\infty]{*} {\cal M}_i(p)$ with :
\begin{itemize}
 \item ${\cal M}_i(p)\in{\cal P}(\mathbb{Z})$,
 \item on the restricted support $[|1,n|]$: ${\cal M}_i(n,p)(dx)={\cal M}_i(p)(dx)$.
\end{itemize}
\end{Thm}

\begin{Thm}[Dirac limit] \label{th2}
We look at the probability ${\cal M}_i(p)(n dx)$ on the space $(\frac{1}{n}\mathbb{N},\mathbb{P}^n)$ where
$\mathbb{P}^n$ puts probability $1$ on points of ${\frac{1}{n}}\mathbb{N}$. As a measure on $\mathbb{R}$, $\frac{1}{n}\mathbb{P}^n$ tends to the
Lebesgue measure on $\mathbb{R}^+$ for weak convergence; we thus have our first scaling: given $p$ and $n\to\infty$: $$\forall p, {\cal
M}_i(p)(ndx)\xrightarrow[n\to\infty]{*} \delta_{0}.$$
\end{Thm}

\begin{Conj}[Fluid limit] \label{th3}
 $n\to\infty$, $p^n=\frac{d}{n}$ consider peer number $i^n=1+\lfloor n \alpha \rfloor$; then there
exists ${\cal M}_{\alpha, d} \in {\cal P}(\mathbb{R})$ that is absolutely continuous with respect to Lebesgue measure such that:
$$\mu^n_{\alpha,d}:={\cal M}_{i^n}(p^n)(n dx)\xrightarrow[n\to\infty]{*} {\cal M}_{\alpha, d}.$$
\end{Conj}

\subsubsection*{Proof of Theorem~\ref{th1}}
Theorem~\ref{th1} is obvious except for the fact that ${\cal M}_i$ has mass $1$; the result essentially comes from the fact that the
probability for peer $i$ to be matched with peer $j$ does not depend on peers with rank greater than the maximum of $i$ and $j$. Thus the
distribution for $n$ peers is only a cut version of the distribution with more peers.

Now we shall prove that the mass is  equal to $1$.
We already know that the mass is equal to $1$ for the exact model (Lemma \ref{Lem:Exact_mass_1}). However, it is not obvious this is still true after the
changes we made to the toy model. The fact that ${\cal M}_i(n,p)\to{\cal M}_i(p)$ gives the mass as an increasing limit. First, suppose the mass
${\cal M}_i(n,p)$ does not tend to $1$. Then there exists some $\epsilon>0$, such that $\sum_{k=1}^{\infty}D(i,k)<1-\epsilon$. If we put this back in
formula (\ref{RecurrenceD}), then:
\begin{equation}\label{eq:conclusion0:masse}
D(i,j)\geq p \epsilon \left(1-\sum_{k=1}^{i-1}D(k,j)\right)
\end{equation}
We know that $(D(i,j))_{j=1..\infty}$ is a sub probability, thus $D(i,j)\to 0$ when $j\to\infty$. From equation (\ref{eq:conclusion0:masse}) it
follows that $\sum_{k=1}^{i-1}D(k,j)\to 1$. A particular consequence is that for $j$ large enough (i.e., there exist $j_0, such that for all j\geq j_0$), we have:
$\sum_{k=1}^{i-1}D(k,j)\geq \frac{1}{2}$ but this is impossible since the $i-1$ sequences $(D(k,j))_{1\leq k < i; j=1..\infty}$ are probabilities.

\subsubsection*{Proof of Theorem~\ref{th2}}
The result is obvious since all the mass stays in compact sets (tightness property on the Polish space $\mathbb{R}$) and ${\cal M}_i(p)$ is a
probability. But the fact is interesting for its physical interpretation.

\subsubsection*{Sketch of proof of Conjecture~\ref{th3}}
This is a very technical result. We will only address here the special case where $\alpha=0$. From a technical point of view, we first have to prove
that the sequence $\mu^n$ is tight, which allows us to extract a limit. We then have to show that this limit is unique.

In the special case $\alpha=0$, let $\beta\in\mathbb{R}$ and $j^n=1+\beta n \lfloor n\rfloor$ then: $D(1,j^n)=p^n\left(1-p^n\right)^{i^n-1}$. This
implies $$nD(1,j^n)\sim d\left(1-\frac{d \beta}{n\beta}\right)^{n\beta} \to d e^{-\beta d}.$$ This in turn yields: $${\cal M}_{0,d}(\mathrm{d}\beta)=d e^{-\beta
d}\mathrm{d}\beta.$$

This theoretical result could be proven though at the expense of very long and technical developments. We do not anticipate any significant 
mathematical difficulty though it does remain to carry through the demonstrations. The results are not necessary to make the following observations, but they explain why we have considered some particular scalings.

\subsection{Observations}\label{subsec:observe}
\begin{figure*}
\centerline{\hspace{-1.4cm} \subfigure[$D(200,j)$]{\includegraphics[width=.38\textwidth]{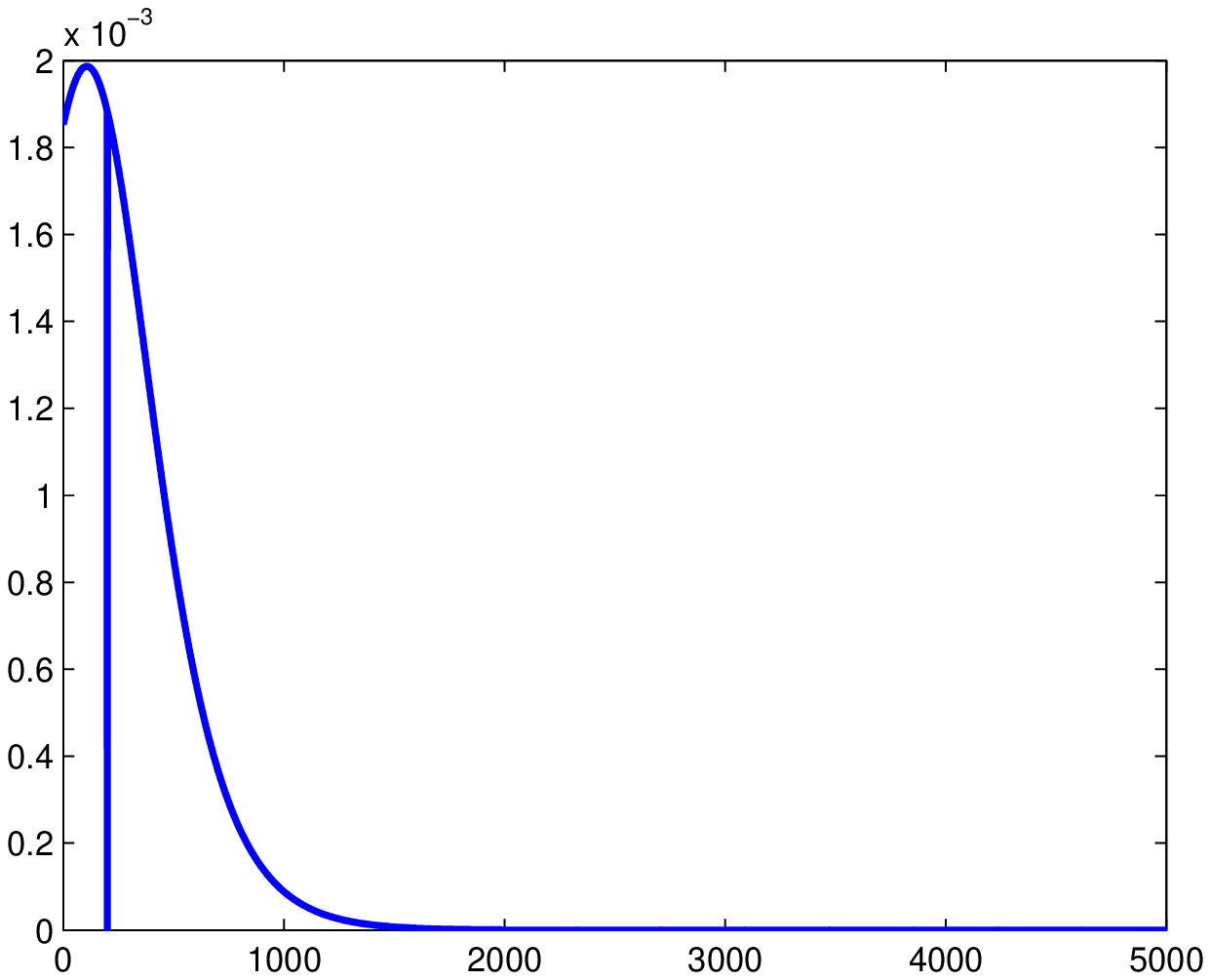}\label{subfig:200}\hspace{-.65cm}}
\subfigure[$D(2500,j)$]{\includegraphics[width=.38\textwidth]{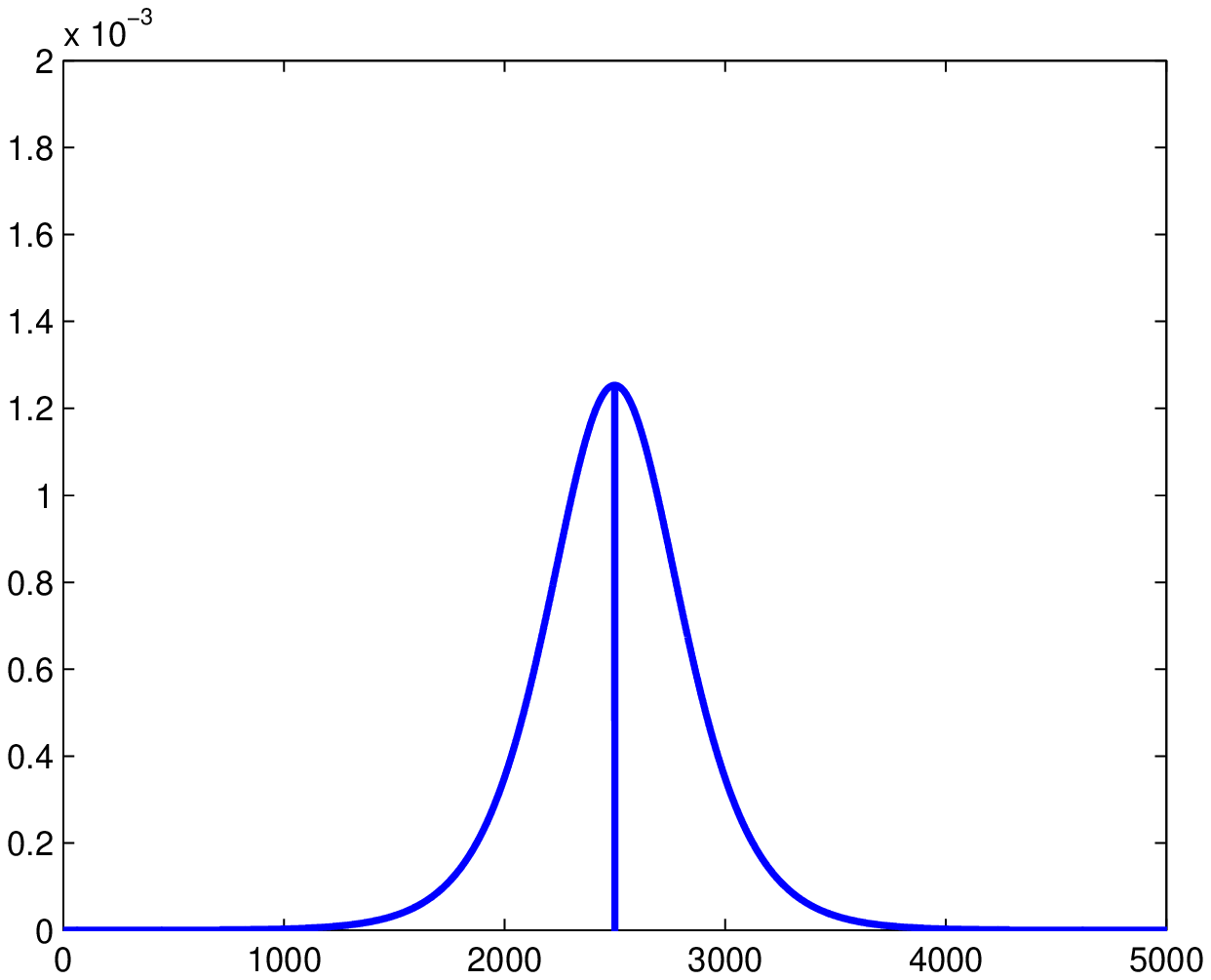}\label{subfig:2500}\hspace{-.65cm}}
\subfigure[$D(4800,j)$]{\includegraphics[width=.38\textwidth]{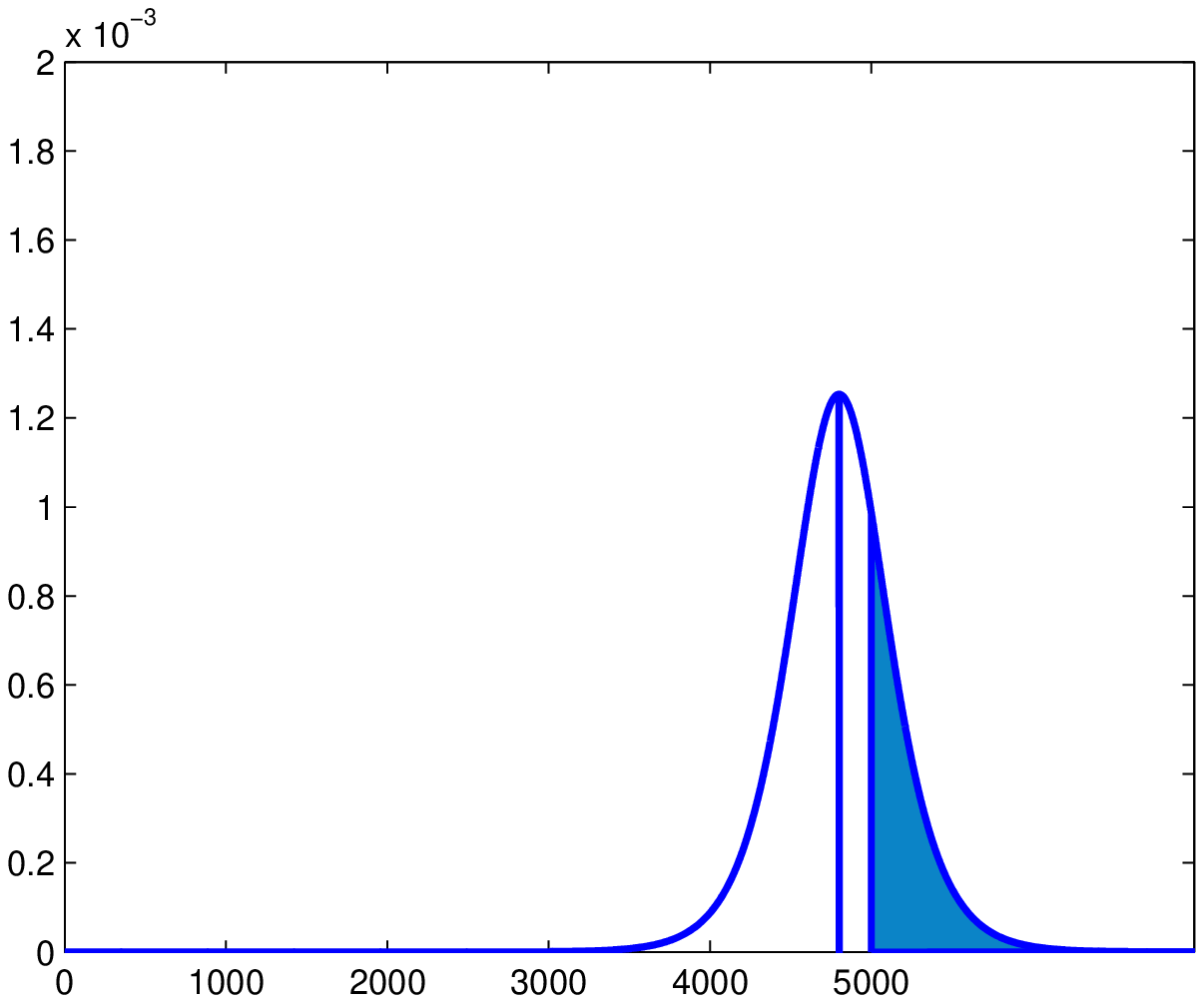}\label{subfig:4800}} \hspace{-1.65cm}} \caption{Distribution of neighbors in
independent $1$-matching for $n=5000$ peers and $p=0.5\%$ for Peer $200$, Peer $2500$ and Peer $4800$.} \label{fig:5000}
\end{figure*}

The results in this section are obtained by solving Equation \ref{RecurrenceD}. We took $n=5000$ to obtain the smoothest possible curves but
$n=100$ would give pretty similar results. In Figure \ref{fig:5000}
we illustrate the different cases that may arise.

In Figure \ref{subfig:200} we see the case of a well ranked peer. Note that for $i=1$ the right part is almost geometrically distributed. Also
note that the best peers are peered with peers of lower average rank, but that this changes quickly and peers in the top $20\%$ but not in
the top $5\%$ have a significantly better mate on average.

The central case is illustrated in Figure \ref{subfig:2500}. We see that the distribution is symmetric and that the
distribution simply shifts with the rank of the peer (for top $25\%$ to top $80\%$ peers). This second fact is a kind of finite horizon property and
illustrates the property we called stratification. Notice that the distribution can not be fit with a normal law, in any case.

In Figure \ref{subfig:4800}, the distribution shift continues for the bottom $20\%$ of peers, but as there is no worse peer to mate with, the
distribution is cut. This means that there is a probability for not being matched which is given by the area filled in blue. A particular case for
the worst peer is that it will be matched exactly in half of the cases. All the others are assured to do better in terms of matching frequency.

\subsection{\texorpdfstring{$b_0$}{b0}-matching independent model}
The $1$-matching case was only presented to give a flavor of the stratification phenomenon. Formally there are no new issues in progressing to a
$b_0$-matching model except for the weight of notation. As for $1$-matching, we state an independence assumption which is not formally true but supplies
a fairly good approximation compared to simulations as shown in paragraph \ref{subsec:validation}.

\subsubsection{Notation}
$n$ still denotes the number of peers. The situation becomes more complicated, because the first choice of one peer may correspond to the last choice of its
mate. Consequently we have to study a quantity $D_{ci}^{cj}(i,j)$ which is not of directly interest. This is the probability that choice
number $ci$ of Peer $i$ is $j$ and that for $j$, $i$ is choice number $cj$. As in the 1-matching case, $D_{ci}^{cj}(i,j)$ does not depend on
larger indexes for $i$, $j$, $ci$ and $cj$. Nor does it depend on $n$. Intuitively this corresponds to the fact that the first choice is made before making the 
second, and that the best peers have priority for choosing their mates. The quantity of interest is
$D_{ci}(i,j)=\sum_{cj=1}^{b_0}D_{ci}^{cj}(i,j)$.

\begin{Ass}\label{b0_AssHole}
Let $i,j\leq n$ and $c_i\geq 1$ and $cj\geq 1$, the events:
\begin{itemize}
\item peer $i$ has chosen $ci-1$ peers better than $j$ and choice $ci$ is not matched by better than $j$,
\item peer $j$ has chosen $cj-1$ peers better than $i$ and choice $cj$ is not matched by better than $i$,
\end{itemize}
are independent.
\end{Ass}

The way to evaluate $D_{ci}^{cj}(i,j)$ is to multiply the probabilities of the supposed independent events:
\begin{itemize}
 \item $i$ knows $j$: with probability $p$,
 \item choice $ci$ of $i$ is not matched and previous choices are matched with better than $j$,
 \item the reciprocal condition on $j$.
\end{itemize}

Note that the probability that choice $ci$ of $i$ is not matched and previous choices are matched with better than $j$ is simply:
$\sum_{k=1}^{i-1}D_{ci-1}(i,k)-\sum_{k=1}^{i-1} D_{ci}(i,k)$, the probability that choice $ci-1$ is matched with better than $j$ minus the
probability that choice $ci$ is matched with better than $j$ (mathematically this formula is exact because one of the two events is included
in the other).

This proves, under assumption \ref{b0_AssHole}, that:
\begin{equation}
D_{ci}^{cj}(i,j)=p\left(\sum_{k=1}^{j-1}D_{cj-1}(j,k)-D_{cj}(j,k)\right) \left(\sum_{k=1}^{i-1}D_{ci-1}(i,k)-D_{ci}(i,k)\right).
\end{equation}

We now show how to compute this formula by recurrence.

\begin{algorithm}[t]\label{alg:independentb}
\setlength{\baselineskip}{0.6\baselineskip} 
\dontprintsemicolon \SetKwData{avc}{cur\_b} \SetKwFunction{connect}{connect} \KwData{ Number of peers, $n$\; Erd\"{o}s-R\'{e}nyi probability, $p$\; Number of
matchings, $b_0$\;} \KwResult{$D_{ci}^{cj}(i,j)$ the probability that the $ci$-th choice of Peer $i$ is $j$ and that the $cj$-th choice of $j$ is $i$
and $D_{c}(i,j)$ the probability that the $c$-th choice of Peer $i$ is $j$\;}
 \BlankLine
$D_{c}\leftarrow zeros(b_0,n,n)$\; $D_{ci}^{cj}\leftarrow zeros(b_0,b_0,n,n)$\; $D_0^c\leftarrow ones(1,b_0,n,n)$\; $D_c^0\leftarrow ones(b_0,1,n,n)$\;

\For{$i = 1$ \emph{\KwTo} $n$} { \For{$j = i+1$ \emph{\KwTo} $n$}{ \For{$(ci,cj) \in [|1,b_0|]\times[|1,b_0|]$
}{\begin{eqnarray}D_{ci}^{cj}(i,j)&\leftarrow&\nonumber p\left(\sum_{k=1}^{j-1}D_{cj-1}(j,k)-D_{cj}(j,k)\right)
\left(\sum_{k=1}^{i-1}D_{ci-1}(i,k)-D_{ci}(i,k)\right)\label{eq:bmatching}\end{eqnarray}\\} \For{$ci = 1$ \emph{\KwTo} $b_0$}{$D_{ci}(i,j)\leftarrow
\sum_{cj=1}^{b_0} D_{ci}^{cj}(i,j)$} \For{$cj = 1$ \emph{\KwTo} $b_0$}{$D_{cj}(j,i)\leftarrow \sum_{ci=1}^{b_0} D_{ci}^{cj}(i,j)$}}\; }
\caption{Independent $b_0$-matching probability computation}
\end{algorithm}

\subsubsection{Independent \texorpdfstring{$b_0$}{b0}-matching algorithm}
Note in the following algorithm that $D_c(i,j)$, the $c$-th choice distribution of $i$ is no longer symmetric for $c>1$, but $D_{ci}^{cj}(i,j)$
has more symmetry (see Algorithm~\ref{alg:independentb}
). Matlab scripts can be found at \cite{Erdos06Matlab_program}. 
This version is not optimized (but sufficiently difficult not to do so); the partial sums can be kept in memory to gain a linear factor.

\subsubsection{Validation of independent \texorpdfstring{$b_0$}{b0}-matching}\label{subsec:validation}
\begin{figure}
\centering
\includegraphics[width=.7\textwidth]{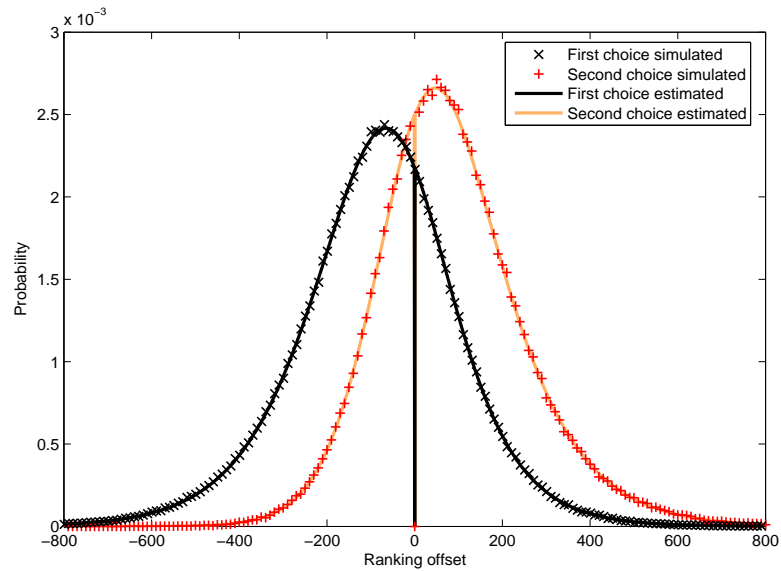}
\caption{Comparison for $n=5000$ and $p=1\%$ (which gives in average $50$ neighbors per peer) of the distributions $D_1$ and $D_2$ for peer $3000$
centered at $3000$ with the statistics obtained by simulating without any approximation many instances of Erd\"{o}s-R\'{e}nyi random graphs.}
\label{fig:3000sur5000_D1_D2_SimulatedvsEstimated}
\end{figure}

As mentioned above, assumptions \ref{AssHole} and \ref{b0_AssHole} work very well except for very small numbers of peers with $p$ very large.
Figure \ref{fig:3000sur5000_D1_D2_SimulatedvsEstimated} illustrates this point. We simulated a $2$-matching by drawing a million realizations of the Erd\"{o}s-R\'{e}nyi graph with $n=5000$ and $p=1\%$ (simulations requiring several weeks) 
and compared  
distributions $D_1(3000,j)$ and $D_2(3000,j)$ with those given by our simplified formula. The comparison in Figure \ref{fig:3000sur5000_D1_D2_SimulatedvsEstimated} illustrates the accuracy of the formula. 

\section{Application to BitTorrent}
\label{sec:consequences}

Results of previous Sections allow us to closely estimate for each peer the ranks of peers it is likely to collaborate with. All our results tend to give a theoretical proof of the stratification phenomenon in systems that use a global ranking function that is not correlated to the acceptance graph. In this Section, we will see how this stratification can give insight about the effect of the Tit-for-Tat policy used in BitTorrent. 

We suppose  that we are in the \emph{post flashcrowd} phase. In the \emph{flashcrowd} phase, an unique seed is uploading a new file, and the upload capacities of the best peers are useless: all peers have downloaded the same blocks. But during the post flash crowd phase, all blocks have roughly the same repartition, because of the download rarest first policy of BitTorrent. So we can assume content availability will not affect the acceptance graph and focus on bandwidth only.

The TFT policy consists in uploading to the peers from which one gets the best download rates. The selection process is renewed periodically. Along with a generous upload connection that allows to probe new peers for an eventual TFT exchange, this acts like the random peer initiative described Section~\ref{sec:context}. This why we claim our results apply to the TFT exchanges in BitTorrent. In peculiar, we have a proof of the stratification effects (peers tend to exchange with peers with similar bandwidths) empirically observed by \cite{bharambe06analyzing,legout06clustering}.

However, the ranking of a peer just gives an intuition about the Quality of Service (QoS) it is presumed to experience. In order to obtain relevant results, it is therefore necessary to bind ranking and performance. In the case of a file sharing system like BitTorrent, the average expected download rate is a very convenient performance metric all the more so since it is easy to compute within our model: it is enough to know
the upload bandwidth for each peer $i$.

\begin{figure}
\centering
\includegraphics[width=.7\textwidth]{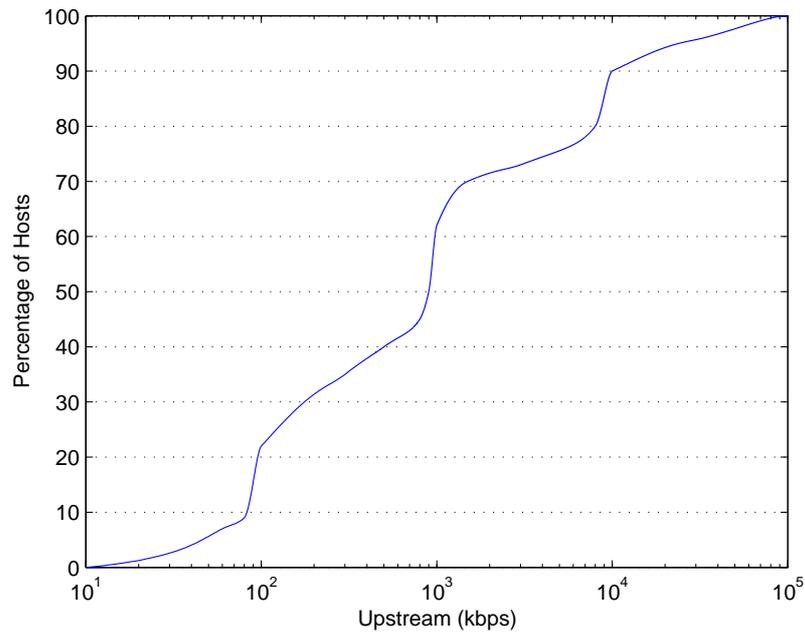}
\caption{Estimation of bandwidth capacities derived from \cite{saroiu02measurement}}
\label{fig:uplink_cdf}
\end{figure}

To compute network performances, we have taken as reference the measurements made by Saroiu \emph{et al.} \cite{saroiu02measurement}. Using bandwidth
estimation in the Gnutella network, they have estimated the upstream for a large community of P2P users. The cumulative distribution they obtained is
shown Figure \ref{fig:uplink_cdf}. One can observe a wide distribution of bandwidths (just like in Orwell's \emph{Animal Farm}, ``all peers are equal but some peers are more equal than others'').

\begin{figure}[t]
\centering
\includegraphics[width=.7\textwidth]{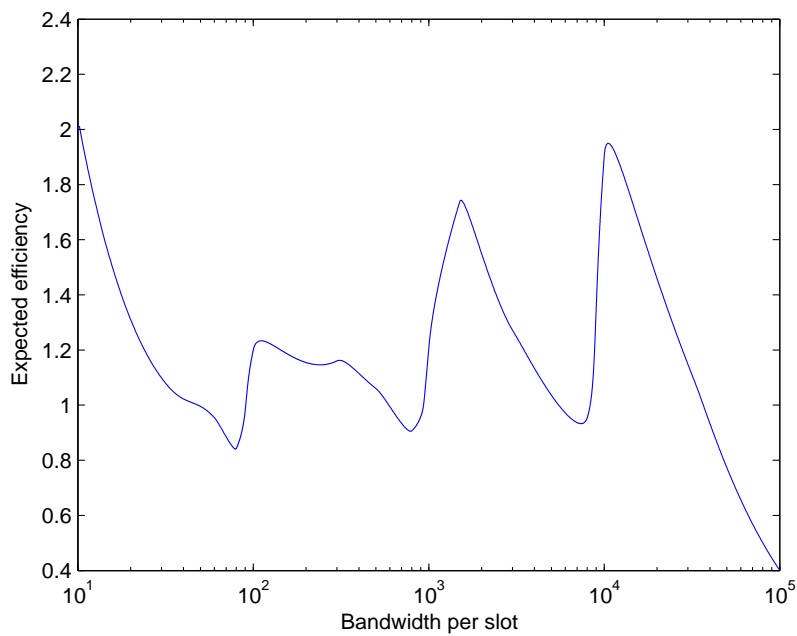}
\caption{Expected D/U ratio as a function of the upload bandwidth offered. $b_0$ is set to 3 and average number of neighbors is $20$.}
\label{fig:20_voisins_ShareRatio}
\end{figure}

Applying our model to the distribution observed by Saroiu \emph{et al.}, we get the results shown in Figure \ref{fig:20_voisins_ShareRatio}.
We chose the following parameters:
\begin{itemize}
\item $b_0$-matching with $b_0=3$, corresponding in a BitTorrent network with all clients having the default number of slots of $4$.
\item expected number of acceptable peers (peers who are known and interesting) $d=20$ (realistic value)
\end{itemize}

\bigskip
Notice that the number $n$ of peers does not have to be given because our model does not depend on the network size: with a partial network
knowledge, observed offsets scale with the number of peers (see Section~\ref{sec:model}).

To put results in the clearest possible way, we chose to represent expected download/upload ratio, which correspond to BitTorrent share ratio. When this
ratio is lesser than $1$, one gives in average more that it receives.

Some observations are worth being said:
\begin{itemize}
\item Best peers suffer from low sharing ratios: as they are the best, they can only collaborate with lower peers, so the exchange is suboptimal for them. The only way for best peers to counter this effect is by adding extra connections until the upload bandwidth per slot is close to the one of lower peers. This somehow explains why BitTorrent proposes \emph{by default} a greater number of connections (up to TCP limitations) for peers with high bandwidths, thus avoiding too much spoil.

\item There is density peeks in the bandwidth distribution. this peeks corresponds to typical Internet connections, such as DSL or cable. Peers in the density peeks have a ratio close to $1$. This is due to the great probability they have to collaborate with peers that have exactly the same characteristics as them.

\item Efficiency peeks appear for peers that have an upload just above a density peek. For these peers, lower peers have almost the same upload bandwidth as them, whereas upper peers are likely to offer greater bandwidth.

\item Surprisingly, the lowest peers have a high efficiency, although there is some probability for them not to be matched, as pointed out by Figure \ref{subfig:4800}. This is related to the relatively high bandwidth (compared to their) they can sometimes obtain: roughly speaking, they can obtain half the time four times their upload bandwidth.
\end{itemize}

As a consequence of this efficiency repartition, it is tempting for an average peer to tweak its number of connection in order to increase the efficiency of its connections. For instance, suppressing one connection can improve the probability of collaborating with higher peers. However, this leads to a Nash equilibrium where all peers have just one TFT slot. This is unacceptable in term of connectivity, but rational peers trying to maximize their benefit cannot be avoided. This is an explanation for the $4$ slots ($3$ TFT and one generous slot) settings: obedient average peers that uses the default settings must have at least $4$ in order to ensure connectivity in the TFT collaboration graph. On the other hand, the more slots they have, the farther they are from the Nash equilibrium that rational peers will try to follow. Hence $4$ seems to be the best trade-off.

\section{Conclusion}
\label{sec:conclusion}

In this paper, we identified the stable matching theory as a natural candidate to model peer-to-peer networks where peers choose their collaborators. Furthermore, we applied elements of this theory to a specific case:  $b$-matching with global rankings. Whereas there has been a lot of work in analyzing incentive to collaborate in some specific application from an economical point of view, this is the first attempt to analyze the behavior of a class of applications using graph theory.

The main conclusion of this study is that matching theory gave insights on the behavior of a  P2P systems class, namely the global ranking class. In both cases of complete and random acceptance graphs, we studied clustering and stratification issues. On most cases, clustering may be prevented using $b$-matching with enough connections and some standard deviation. But stratification is an intrinsic property of such networks. It seems impossible to overcome it as long as each peer follows the \emph{try-to-collaborate-with-the-best} rule. Interestingly, for random overlay graphs, the crucial parameter is $d$, the average number of acceptable peers, which makes stratification a flawlessly scalable phenomenon.

As a first application, our results provide some new insights on BitTorrent parameters. They show that best peers have to set up a large number of connections in order to avoid bad download/upload ratio. The \emph{by default} number of collaboration ($4$) is justified. It allows, to a certain extend, to maintain connectivity in the TFT exchanges and  to protect peers using default settings (obedient peers) from peers with optimized settings (rational peers).

When considering the stable properties which emerge, it also become clear that different class of utility functions leads to very different properties. This can be exploited according to the needs of the targeted application. For example, in a peer-to-peer streaming protocol, the most important feature is a small play out delay but a strong stratification, needed to give peers incentive to collaborate, produce a collaboration graph with large diameter (large play out delay). In many cases, combining different utility function will be necessary. Such a combination can, for instance, be achieved by introducing a second type of collaborations depending on a different global ranking or depending on a symmetric ranking such as latency.

\textsl{\textbf{Acknowledgment:} The authors wish to thank James Roberts and Dmitri Lebedev for their helpful comments}

\bibliographystyle{plain}
\bibliography{maria}

\end{document}